# The calculation of the effective interaction parameter in LDA+U method by linear response approach for Fe(OH)$_2$

Mansoureh Pashangpour [1], Khadijeh Imani [2], Amir Abbas Sabouri-Dodaran [3,4], Nasser Nafari [3]

[1] *Plasma Physics Research Center, Science and Research Campus, Islamic Azad University, P.O.Box 14665-678, Tehran, Iran*
[2] *Islamic Azad University, Karaj Branch, Tehran, Iran*
[3] *Institute for Theoretical Physics and Mathematics, Tehran, 19395-5531, Iran*
[4] *Payame Noor University, Tehran. 19395-4697, Iran*

In this paper we have investigated the electronic properties of Fe(OH)$_2$ hydroxide by using the LSDA+U as well as the generalized gradient approximation. Our calculations for the iron-hydroxide show that the LSDA results are greatly at variance with experimental findings. On the other hand we have shown that LSDA+U is capable of opening a gap at the Fermi level resulting in insulating ground state in agreement with experimental observations.

## I. INTRODUCTION

Several authors have investigated the electronic properties of layered hydroxides[1]. In particular, brucite type hydroxides with generalized formula M$^{2+}$(OH)$_2$, where M$^{2+}$ stands for a divalent cation with layered structure is studied extensively. In this paper we have been working on the hydroxide Fe(OH)$_2$, commonly known as white rust. Fe(OH)$_2$ has a hexagonal crystal and is isostructural with brucite, Mg(OH)$_2$, and the simple hydroxides Ca(OH)$_2$, Mn(OH)$_2$, β-Co(OH)$_2$ and β-Ni(OH)$_2$. Factor group analysis at the Brillouin zone center determines that there are two internal stretching OH vibrations, i.e., the symmetric A$_{1g}$(OH) and the antisymmetric A$_{2u}$(OH) modes, and six lattice modes A$_{1g}$(T), A$_{2u}$(T), E$_g$(T), E$_u$(T), E$_g$(R), E$_u$(R)[2].

Both the local spin-density approximation (LSDA) and the spin polarized version of the generalized gradient approximation (σ-GGA) fail to predict the insulating behavior of Fe(OH)$_2$ and many simple transition metal oxides (TMO). The LDA+U approach, first introduced by Anisimov et al.[3, 9-12], has allowed the study of a large variety of strongly correlated systems with considerable improvement with respect to LSDA or σ-GGA results. In fact, the standard density functional theory (DFT) which employs LDA or GGA, produce an unphysical metallic character due to the fact that crystal field and electronic structure effects are not sufficient in this case to open a gap. The use of the LSDA+U method for studying Fe(OH)$_2$ is mainly motivated by the attempt to reproduce the observed insulating behavior.

The organization of this paper is as follows: In section II we present the LSDA+U approach which starts with the formulation of Anisimov et al.[4-6]. In section III we use a method, based on linear response approach to calculate the interaction parameters entering the LSDA+U functional, in an internally consistent way[8]. Finally, in sections IV and V, we apply this methodology to study the electronic properties of Fe(OH)$_2$ as a strongly correlated system.

## II. STANDARD LDA+U IMPLEMENTATION

Anisimov *et al.* have extended the standard energy functional in LDA to LDA+U by adding an on-site Hubbard-like interaction E$_{Hub}$, in order to account for the on-site Coulomb interaction responsible for the correlation gap in Mott insulators[4-6]:

$$E_{LDA+U}[n(\vec{r})] = E_{LDA}[n(\vec{r})] + E_{Hub}[\{n_m^{I\sigma}\}] - E_{DC}[\{n^{I\sigma}\}],$$

where $n(\vec{r})$ is the electronic density and $n_m^{I\sigma}$ are the atomic orbital occupations for the atom I with spin σ. The last term in this equation is to avoid double counting of the interactions contained both in the

$E_{Hub}$ and, in some averaged way, in $E_{LDA}$. In this term $n^{I\sigma}$ is the trace of $n_m^{I\sigma}$ over magnetic quantum number $m$. A rotationally invariant formulation has been introduced[7,9] where the orbital dependence of $E_{Hub}$ is taken from the atomic Hartree-Fock with renormalized Slater integrals.

## III. RESPONSE FUNCTION METHOD

In this procedure we introduce the response functions as the localized level occupations with respect to the potential shift in $d$-levels. By using the response function approach[8,10], the effective interaction parameter $U$ is calculated in the following equations:

$$\chi_{IJ} = \frac{dn_d^I}{d\alpha_J}, \quad (1)$$

$$U = -\frac{d\alpha^I}{dn_d^I} + \frac{d\alpha^I}{dn_{d0}^I} = \left(\chi_0^{-1} - \chi^{-1}\right)_{II}. \quad (2)$$

Here $\chi_{IJ}$ are the elements of the linear response matrix, $n_d^I$ is the occupation number of $d$-levels and $\alpha^I$ is the potential shift in $d$-levels. $\chi$ (the screened response matrix) includes all screening effect from crystal environment which is associated with the localized electrons and $\chi^0$ (the unscreened response matrix) contains non-screening effect of the total energy of the non-interacting Kohn-Sham associated with the system. $\chi^0$ must be subtracted from $\chi$ to eliminate these effects for evaluation of the physical value of the Hubbard $U$. To compute the Hubbard effective interaction $U$, we used LSDA calculations with potential shift acting on the $d$-levels in one of the Hubbard atoms which in this case are the Iron sites. For evaluating the on-site coulomb interaction $U$, we only consider the nearest neighbor electronic interactions of Iron-sites and in our numerical calculations we use different number of iron atoms in a given supercell. We perform a series of perturbations on Hubbard atoms (iron atoms). Moreover, we consider a delocalized background which adds one more column and row to the response matrix. These elements are determined by imposing overall charge neutrality of the perturbed system for all localized perturbations:

$$\sum_I \chi_{IJ} = 0, \quad \sum_I \chi_{IJ}^0 = 0 \text{ for all } J, \quad \sum_J \chi_{IJ} = 0,$$
$$\sum_J \chi_{IJ}^0 = 0 \text{ for all } I. \quad (3)$$

The singularities in $\chi^{-1}$ and $\chi_0^{-1}$ cancel out when computing the difference $\chi_0^{-1} - \chi^{-1}$ by shifting the elements of the $\chi, \chi^0$ matrices with the same amount.

## IV. COMPUTATIONAL DETAILS

Our calculations are based on the use of DFT and the *ab initio* pseudo-potential plane-wave method using the PWSCF code of the Quantum ESPRESSO distribution[11]. We start our calculations with a hexagonal unit cell (space group $p\bar{3}m1$) and assume that the hydroxide Fe(OH)$_2$ is in antiferromagnetic with the magnetic moments aligned in the basal plane. Fe atoms are located at (0, 0, 0), (0, 0, 0.5). The O and H atoms are at (1/3, 2/3, z), (1/3, 2/3, z+0.5), (2/3, 1/3, 0.5-z), (2/3, 1/3, 1-z) with z ~ 0.1, for O and z ~ 0.2 for H atoms. All coordinate numbers are specified in crystal unit. Experimental lattice parameters are a= 6.162 a.u. and c= 17.832 a.u.[12].

The calculations are performed with ultrasoft GGA Perdew-Burke-Ernzherof[13] (US PBE) nonlinear core-correction (NLCC) pseudopotential for iron atoms while for Oxygen and Hydrogen US PBE (non NLCC) potentials have been chosen. We also perform the calculations with US LDA Perdew-Zunger[14] (PZ) NLCC for iron atoms and US PZ (non NLCC) potentials for oxygen and Hydrogen. For the pseudopotential generation we use the $3d^7 4s^1$ valence-atomic-configuration for Iron, the $2s^2 2p^4$ for Oxygen and the $1s^1$ for Hydrogen. Brillouin Zone integrations were performed using $6\times6\times3$ Monkhorst and Pack[15] special point grids using Marzari and Vanderbilt smearing technique with a smearing width of 0.005 Ry in order to smooth the Fermi distribution. The Kohn-Sham orbitals are expanded in a plane wave basis set with a 50 Ry energy cutoff and 400 Ry for the charge density due to the use of ultrasoft pseudopotentials for Fe, O and H and periodic boundary conditions are imposed. The optimized lattice parameters of Fe(OH)$_2$ after relaxation are a= 6.152 a.u. , c= 17.787 a.u. and z = 0.109 for O atoms, z= 0.214 for H atoms in crystal unit. So the optimized volume is 0.5% less than the experimental. In order to compute the Hubbard interaction $U$, we use potential shifting which acts on the iron ion to study the response of the $d$ atomic occupations on the perturbed site and on the other atoms in the system. For this purpose, we perform the series of self consistent calculations with different potential shifts in the $d$-levels of one of the iron ions. Then, through the equations (1) and (2), we obtain the response matrices $\chi, \chi^0$. Practically, the perturbed atom strongly interacts with its nearest neighbors with the same spin being translationally equivalent to it, also it has a shifted potential in the $d$ channel, so that screening process is not completely efficient. Thus, we perform perturbation in a larger supercell. First, we consider C1 supercell with two iron ions and then extrapolate the result to C4 supercell which contains four C1 supercells with eight iron ions (Fig. 1).

## V. LINEAR RESPONSE IN SUPERCELLS

### A. C1 supercell

The C1 supercell contains two iron atoms in a hexagonal structure with opposite spin polarizations (see Fe1 and Fe2 in Fig.(1a)).
Fig.2 shows the variation of $d$-level occupation of Fe ions in terms of potential shifts in the first iteration of perturbation (the density response which is obtained at the first iteration does not involve any effect of the electron interaction and actually corresponds to the response of the independent electron system) when the perturbed atom is Fe1, which these define the non screening part of linear response $\chi^0$. The slopes of the two lines in Fig. 2 are -1.2464 Ry$^{-1}$ and 1.0634 Ry$^{-1}$, but if the perturbed atom is Fe2, these slopes are slightly shifted relative to the previous result. In this way we obtain the elements of the non-screening linear response matrix, i.e. , $\chi^0_{11}, \chi^0_{12}, \chi^0_{21}, \chi^0_{22}$.

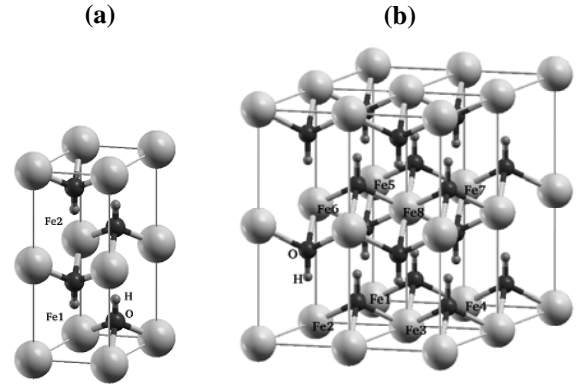

Fig. 1. (a) C1 supercell with opposite spin polarization for the two iron atoms, Fe1 and Fe2 (b) C4 supercell with 8 iron atoms.

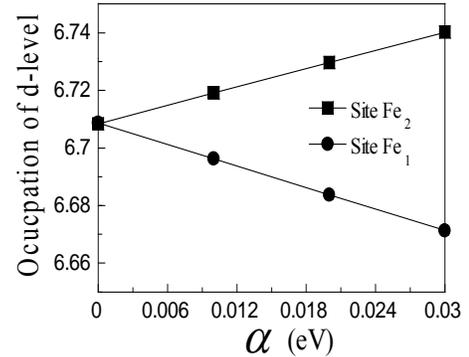

**Fig. 2.** Occupation of $d$-levels of Fe ions in terms of potential shifts for Fe ions from first iteration of perturbation, i. e., independent electron system response, when Fe1 in supercell C1 is perturbed.

In the second step, the information of the last iteration of perturbation, when electron-electron interaction has fully played their role in screening the effective coupling and charge distributions, gives us the screening linear response matrix $\chi$. Fig. 3 shows the variations of $d$-levels occupation of Fe ions in terms of potential shifts in the last iteration of perturbation when the perturbed atom is Fe1. The slopes of the lines in Fig. 3 are -0.14601 Ry$^{-1}$ and 0.018148 Ry$^{-1}$. We obtain a similar figure when the perturbed atom is Fe2. In this case the slopes are 0.017886 Ry$^{-1}$ and -0.14572 Ry$^{-1}$.

levels of one particular iron ion in the C4 supercell (The results are shown in Fig. 4). According to Fig. (4.a), the slopes of the lines are 0.51293 Ry$^{-1}$, -2.998 Ry$^{-1}$, 0.3246 Ry$^{-1}$, and 0.32439 Ry$^{-1}$ and the value of these slopes in Fig. (4.b), when the perturbed ion is Fe$_2$ in C4 supercell, are 0.016778 Ry$^{-1}$, -0.1949 Ry$^{-1}$, 0.0047936 Ry$^{-1}$ and 0.005706 Ry$^{-1}$.

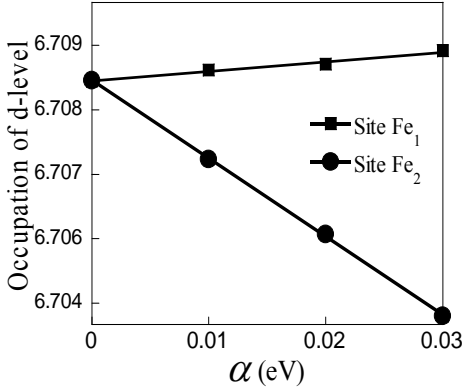

Fig. 3. Occupation of $d$-levels of Fe ions in terms of potential shifts on Fe1 ion in the final iteration of perturbation, i. e., fully screening response, when Fe1 in supercell C1 is perturbed.

Considering equation (3), this can be easily done by adding the delocalized background to both $\chi, \chi^0$. Diagonal elements of $\chi_0^{-1} - \chi^{-1}$ show the value of the effective interaction parameter for C1 supercell as 2.96 eV and non-diagonal terms are corresponded to the inter-site effective interaction in the local density approximation.

## B. C4 supercell

The result obtained using the C1 supercell can be corrected by considering a larger supercell, C4 supercell, with 8 iron ions in the unit cell (Fig. (1b)).

By considering the nearest neighbors contribution, the effective interaction can be estimated most efficiently. Thus in order to obtain the response functions we first perturb the potential on the $d$-

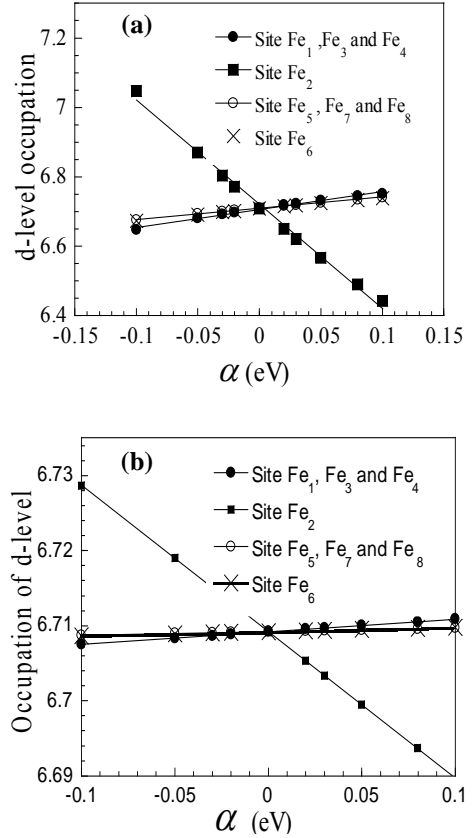

**Fig. 4.** Occupation of $d$-levels of irons in terms of potential shifts on Fe2 ion for C4 supercell (a) in the first iteration (non screening part of linear response, $\chi^0$) (b) in the final iteration (fully screening part of linear response, $\chi$).

By considering the background effects in the $\chi$ and $\chi_0$ response matrices, the diagonal elements of the matrix $\chi_0^{-1} - \chi^{-1}$ gives us the value of the Hubbard $U$ on the iron sites in C4 supercell. The result is $U$= 4.05 eV. Then, we have used this computed interaction parameter in our LDA+U calculations to obtain some of the physical properties of this compound. Fig. 5 demonstrates the change of the total density of states of bulk Fe(OH)$_2$ considering the on-site effective Coulomb interaction $U$ in the two cases of 0 and 4.05 eV. By taking into account the on-site correlation effects, Fe-$d$ bands below/above the Fermi level are shifted to the lower /higher energy level, which results in a notable gap.

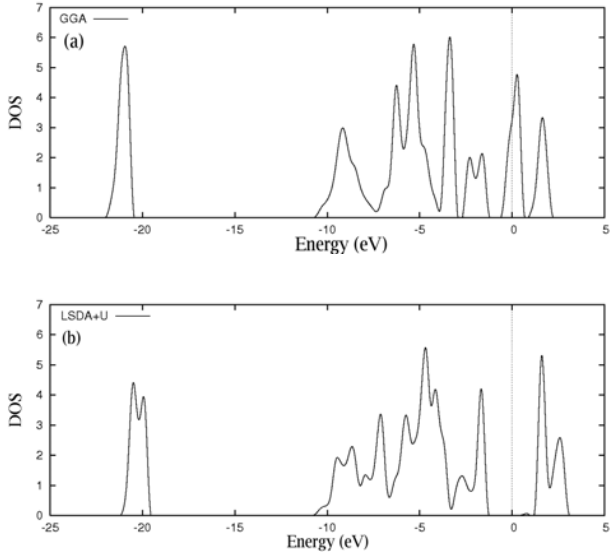

**Fig. 5.** Density of states of Fe(OH)$_2$ as a function of energy with (a) GGA and (b) LSDA+U ($U$=4.05 eV).

Apart from the change of density of states (DOS), one can observe that there are changes of the projected density of states (PDOS) weights near the Fermi level (Fig.6). The minority-spin (down states of iron) $t_{2g}$ manifold of iron ions, that within GGA crosses the Fermi energy, is split into two subgroups by the gap opening. The lower energy minority-spin $d$ states combine in the group of states below the Fermi level where they mix strongly with the states originating from oxygen $p$ orbital. The $d$-states of the iron atom are split to $t_{2g}$- and $e_g$-states; the five $d$-electrons occupy the majority spin (up states of iron), and the remaining $d$-electron occupies one state of the minority spin.

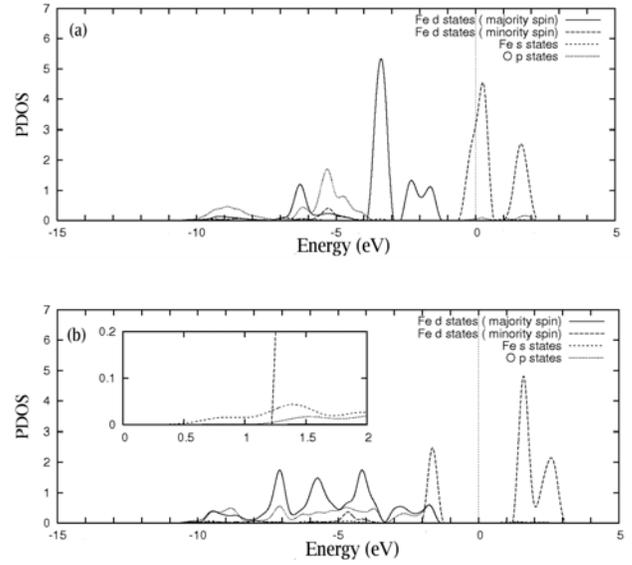

**Fig. 6.** The atomic-projected density of states of Fe(OH)$_2$ as a function of energy with (a) GGA and (b) LSDA+U ($U$=4.05 eV).

The calculated band structure of Fe(OH)$_2$ with $U$=4.05 eV is shown in Fig. (7b). We obtain the insulating behavior as shown in the band structure plot of Fig .(7a) where a comparison is made with GGA (metallic) results. At variance with the GGA results a band gap of about 2 eV now separates the valence manifold from the conduction one. The band structure width around Fermi level (W) in Fig. (7a) is 0.9± 0.1 eV so we obtain the ratio of U/W about 4.5± 0.5 which shows this material can be considered as a strongly correlated system. The gap opens around the Fermi level with a minimal width of about 2 eV. The band gap is direct and located at the Γ point.

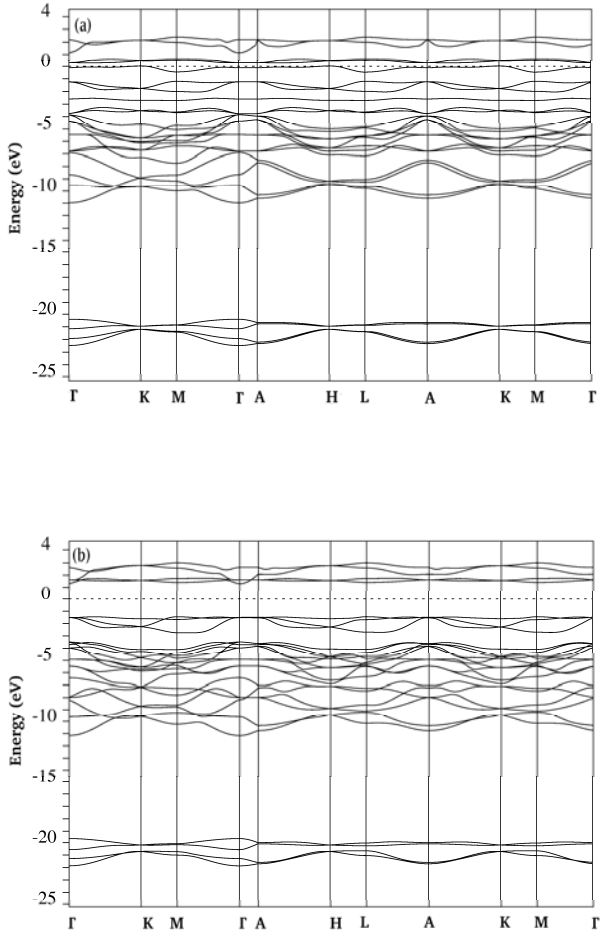

**Fig. 7.** The band structure for Fe(OH)2 with (a) GGA and (b) LSDA+U ($U$=4.05 eV).

A weak transition correspond with {$3d$(Fe), $2p$(O)} → $4s$(Fe) will be predicted around 2 eV due to the vanishing weight of $s$ states of iron ions at the bottom of the valence band (Fig. (6b)). Also a stronger absorption line will be observed around 2.4 eV which corresponds with the transition of {$3d$(Fe), $2p$(O)} → $3d$(Fe) character between two peaks of the density of states around the Fermi level.

## VI. CONCLUSIONS

In this work we have used LSDA+U formulation and a developed method that is based on a linear response approach to calculate the interaction parameter (Hubbard $U$) entering the LSDA+U functional for Fe(OH)$_2$. By this approach we obtained $U$=4.05 eV. The calculated band structure of this compound within the LDA+U approach shows a gap around the Fermi level whose minimal width is about 2 eV. The band gap is direct and located at the Γ point. It corresponds to a weak transition of {$3d$(Fe), $2p$(O)} → $4s$(Fe) character. The stronger transition of {$3d$(Fe), $2p$(O)} → $3d$(Fe) character appears around 2.4 eV. Using the band structures within GGA and LSDA+U shows the ratio of U/W is 4.5± 0.5. In both calculations, LSDA+U and GGA, the magnetic moment of Fe$^{2+}$ in Fe(OH)$_2$ is 3.98 $\mu_B$.

## ACKNOWLEDGMENT

This work has been supported by the computational Nanotechnology Supercomputing Centre Institute for Research in fundamental Sciences (IPM) Tehran, Iran. We would like to thank S. Scandolo and Stefano de Gironcoli for their valuable discussions.